%% file: arxiv.tex
\documentclass[reprint,amsmath,amssymb, aps,pre,floatfix,superscriptaddress]{revtex4-2}
\usepackage{graphicx} 
\usepackage{amsmath}
\usepackage{subfigure}
\usepackage{physics}
\usepackage{appendix}
\usepackage{makecell}
\usepackage{natbib}
\usepackage{graphicx} 
\usepackage{siunitx}
\usepackage[version=4]{mhchem}
\usepackage{comment}
\usepackage{breqn}
\usepackage{xcolor}
\usepackage{tikz}
\usepackage{scalerel}
\usepackage{xifthen}
\usepackage{chngcntr}
\usepackage{xcolor}
\usepackage{bm}
\usepackage{physics}
\usepackage[utf8]{inputenc}
\usepackage{tcolorbox}
\usepackage{mdframed}
\usepackage[T1]{fontenc}
\usepackage{amsmath}
\usepackage{amsfonts}
\usepackage{amssymb}
\usepackage{algorithm}
\usepackage{algpseudocode}
\usepackage{float} 
\usepackage{array}
\usepackage{makecell}
\usepackage{multirow}

\usepackage{outlines}
\usepackage{colortbl}
\usepackage{booktabs}
\usepackage[bottom]{footmisc}
\usepackage{tcolorbox}
\usepackage{diagbox}
\usepackage{bbding}
\usepackage{pifont}
\usepackage{wasysym}

\usepackage{titlesec}

\usepackage{listings}
\usepackage{xcolor}

\lstdefinestyle{mypython}{
    language=Python,
    backgroundcolor=\color{gray!20}, 
    basicstyle=\ttfamily\small,
    keywordstyle=\color{blue},
    commentstyle=\color{green!50!black},
    stringstyle=\color{orange},
    showstringspaces=false,
    breaklines=true
}

\usetikzlibrary{svg.path}
\usepackage[colorlinks,allcolors=blue]{hyperref}

\definecolor{orcidlogocol}{HTML}{A6CE39}
\tikzset{
  orcidlogo/.pic={
    \fill[orcidlogocol] svg{M256,128c0,70.7-57.3,128-128,128C57.3,256,0,198.7,0,128C0,57.3,57.3,0,128,0C198.7,0,256,57.3,256,128z};
    \fill[white] svg{M86.3,186.2H70.9V79.1h15.4v48.4V186.2z}
 svg{M108.9,79.1h41.6c39.6,0,57,28.3,57,53.6c0,27.5-21.5,53.6-56.8,53.6h-41.8V79.1z M124.3,172.4h24.5c34.9,0,42.9-26.5,42.9-39.7c0-21.5-13.7-39.7-43.7-39.7h-23.7V172.4z}
 svg{M88.7,56.8c0,5.5-4.5,10.1-10.1,10.1c-5.6,0-10.1-4.6-10.1-10.1c0-5.6,4.5-10.1,10.1-10.1C84.2,46.7,88.7,51.3,88.7,56.8z};
  }
}

\newcommand\orcid[1]{\href{https://orcid.org/#1}{\mbox{\scalerel*{
\begin{tikzpicture}[yscale=-1,transform shape]
\pic{orcidlogo};
\end{tikzpicture}
}{|}}}}

\newcommand{\nus}{Department of Mechanical Engineering, National University of Singapore, 117575, Singapore}

\input{sym_math}

\input{sym}

\begin{document}

\title{Automated Code Development for PDE Solvers Using Large Language Models}

\author{Haoyang Wu~\orcid{0009-0008-1631-2785}}
\affiliation{\nus}
\author{Xinxin Zhang~\orcid{0000-0003-0297-7061}}
\affiliation{\nus}
\author{Lailai Zhu~\orcid{0000-0002-3443-0709}}
\email{lailai\_zhu@nus.edu.sg}

\begin{abstract}
Foundation models---large language models (LLMs) in particular---have become ubiquitous, shaping daily life and driving breakthroughs across science, engineering, and technology. Harnessing their broad cross-domain knowledge, text-processing, and reasoning abilities for software development, \eg numerical libraries for solving partial differential equations (PDEs), is therefore attracting growing interest. 
Yet existing studies mainly automate case setup and execution for end users. We introduce \pde, a zero-shot, multi-agent LLM framework that automates code development for PDE libraries, specifically targeting  secondary developers. By translating mathematical and algorithmic descriptions directly into source code, \pde ~generates new solvers/modules and adapts existing ones. This end-to-end math-to-code approach enables a self-augmenting pipeline that continuously expands the codebase of a library, extends its capacities, and broadens its scope. We demonstrate \pde ~on three tasks: 1) build a solver for a new PDE, 2) implement new BCs for a given PDE, and 3) modify an existing solver to incorporate additional terms, achieving moderate success rates. 
Failures due to syntactic errors made by LLMs are analyzed and we propose effective fixes. We also identify the mechanisms underlying certain semantic errors, guiding future research.
\end{abstract}

\maketitle

\section{Introduction}
The evolution of computer programming is deeply intertwined with the history of human ingenuity in devising tools to communicate instructions to machines. Since the advent of modern electrically powered computers in the 1940s, programming---specifically coding---probably represents one of the most intellectually intensive human activities. 
Writing code is typically a time-consuming, cognitively demanding, and error-prone manual process, necessitating tedious and painstaking debugging. However, with the recent advances in generative artificial intelligence (AI), there are emerging signs of relief for code developers, offering a promising glimpse into more efficient coding paradigms. Indeed, the widespread adoption of generative AI technologies, such as large language models (LLMs), has demonstrated their potential to fundamentally transform how work is performed in diverse areas including, to name a few, image synthesis~\cite{rombach2022high}, video generation~\cite{brooks2024video}, mesh generation~\cite{xu2023hierarchical}, predictions of protein structure~\cite{jumper2021highly}, 
robotic control~\cite{zeng2023large, wang2024prompt,xu2025training,kopitca2025application}, engineering optimization and design~\cite{rios2023large,zhang2025using,jiang2025generative},
and coding~\cite{liu2023your,jiang2024survey}, etc.

Notably, LLMs have reshaped the landscape of code development and software engineering more broadly, where LLM-driven code generation stands as a particularly fascinating and impactful application. This new paradigm of code generation refers to translating natural language descriptions into source code, which has been employed to develop research prototypes for various applications such as web development~\cite{calo2023leveraging,toth2024llms},  chip design~\cite{thakur2023benchmarking,blocklove2023chip,lu2024rtllm}, robotic simulation~\cite{wang2023gensim}, mathematical theorem proving~\cite{wang2024theoremllama}, computer-aided design~\cite{badagabettu2024query2cad}, and database interface via SQL~\cite{hong2024next}.

Besides these applications, another promising avenue is harnessing LLMs to automate the workflow of solving partial differential equations (PDEs)~\footnote{We regard ordinary differential equations (ODEs) as reduced PDEs.} or to automate software development  for PDE solvers and libraries. 
PDEs are not only ubiquitous in physics and engineering but also find extensive applications in other domains including biology, finance, computer science, the social sciences, and so on. Naturally, it is compelling to contemplate: Can the combination of LLMs' extensive knowledge, code-as-text processing, and reasoning ability,  
spark a breakthrough
automatic generation of PDE-solver modules  directly from plain-language and/or mathematical problem descriptions?  
Indeed, recent works spearheading this direction have revealed the potential of LLMs paving the way towards this visionary objective, as evidenced below.

\begin{table*}[]
\begin{tabular}{c|c|c|c|c|c|c|c}
 Studies & \thead{Published \\date} & LLM & \thead{Prompt or \\ API agent(s)}  & \thead{External \\ package} & \thead{Spatial \\ discretization} & \thead{Generate  \\ new code}    \\
\hline\hline
 \cite{kashefi2023chatgpt}& 2023-03-21 & ChatGPT & Prompt &\thead{NumPy; \\ MATLAB}  & \thead{FDM;\\FVM \\ (Godunov \\method)}  & \checkmark   \\
 \hline
 \cite{ai4science2023impact}& 2023-11-13 & GPT-4  & Prompt & MATLAB &FDM & \checkmark   \\
 \hline
\cite{ni2024mechagents}& 2023-11-14 & \thead{GPT-4 \\ (gpt-4-0613) }& \thead{API \\ agents}& FEniCS &  FEM & \checkmark    \\
 \hline
 \cite{ali2024physics}& 2023-12-04 & \thead{GPT-4 \\ (gpt-4-0314)} & \thead{A single \\ API agent} & Dedalus & \thead{Spectral \\ method} & \checkmark   \\
 \hline
\cite{kim2024chatgpt}& 2024-03-29 & GPT-4 & Prompt & MATLAB  & FDM & \checkmark    \\
 \hline
\cite{chen2024metaopenfoam}& 2024-07-31  & GPT-4o & \thead{API \\ agents} & OpenFOAM & FVM & \ding{55}    \\
 \hline
 \cite{tian2024optimizing}& 2024-08-23  & GPT-3.5 Turbo & \thead{API \\ agents} & FEniCS & FEM & \checkmark   \\
  \hline
 \cite{mudur2024feabench} & 2024-09-28 & \thead{Claude 3.5 Sonnet; \\GPT-4o; \\ Gemini-1.5-Pro; \\ CodeGemma-7B-IT; \\ Gemma-2-27B-IT; \\ Gemma-2-9B-IT} & \thead{A single\\ API agent} & \thead{COMSOL \\ Multiphysics}
 & FEM & \ding{55}   \\
 \hline
 \cite{pandey2025openfoamgpt} & 2025-01-10 & \thead{Claude 3.5 Sonnet \\ o1-preview \\ Gemini-1.5-Pro} & \thead{A single \\ API agent} & \thead{OpenFOAM}
 & FVM & \ding{55}   \\ 
 \hline
 \cite{elrefaie2025ai} & 2025-03-30 & GPT-3.5 Turbo  & \thead{API\\ agents} & \thead{OpenFOAM}
 & FVM & \ding{55}  \\
  \hline
  
 \cite{zhang2025mooseagent} & 2025-04-11 & DeepSeek-R1 and V3 & \thead{API \\ agents} & MOOSE
 & FEM & \ding{55}  \\
 \hline 
 \cite{feng2025openfoamgpt} & 2025-04-27 & Claude 3.7 Sonnet & \thead{API \\ agents} & \thead{OpenFOAM}
 & FVM & \ding{55}  \\
\hline
 \cite{xu2025cfdagent} & 2025-07-31  & \thead{GPT-4o \\ GPT-4o-mini} & \thead{API \\ agents}   &  \thead{In-house \\ solver} & FVM  & \ding{55} \\
 \hline
\thead{Current \\ study} & 2025-08-08 & \thead{Claude 3.5 Sonnet \\ o1-preview \\ o3-mini} & \thead{API \\ agents}   & XLB & LBM  & \checkmark   
\end{tabular}
\label{tab:pdesolver}
\caption{
Related works ordered by publication date.
}
\end{table*}

\section{Related Works}\label{sec:related}

\cite{kashefi2023chatgpt} pioneered the use of prompting LLMs, ChatGPT specifically, to generate code for addressing diverse numerical problems and machine learning settings; the former include solving various PDEs, \eg Poisson equation, diffusion equation, and incompressible Navier-Stokes (NS) 
equations in two dimensions, among others. 

In testing the one-dimensional porous medium equation, \cite{ai4science2023impact} prompts GPT-4 to write a MATLAB solver and to compare the numerical solution with the provided analytical counterpart. Despite multiple attempts, the generated code based on a finite-difference method (FDM) still fails to yield correct solutions due to an oversight of the time step constraint.

\cite{ni2024mechagents} develops a multi-agent framework based on OpenAI's GPT-4 API, which can generate, execute, and correct code using the open-source computing platform for solving PDEs based on finite element method (FEM), FEniCS, to solve classical elasticity problems.

By testing four well-documented and widely used scientific packages, \cite{ali2024physics} evaluates the coding abilities of the OpenAI API (gpt-4-0314) for physics simulations. When prompted to use Dedalus, an open-source Python PDE solver based on spectral methods, the LLM generates code to solve the one-dimensional diffusion equation, scalar advection equation, and nonlinear acoustic wave equation, among others. However, the pass rate of the generated code is below $50\%$.

\cite{kim2024chatgpt} employs ChatGPT-4 via prompts to generate MATLAB code for simulating a two-dimensional seepage flow, specifically solving a Laplace equation based on a central FDM. To generate functional code that yields correct solutions, manual iterations of error checking with subsequent prompt refinement are necessitated.

\cite{chen2024metaopenfoam} introduces MetaOpenFOAM, a natural language-based automation framework designed to perform computational fluid dynamics (CFD) simulations.
This framework drives OpenFOAM---a PDE solver that employs the finite-volume method (FVM)---to solve the NS equations. 
Leveraging MetaGPT~\cite{hong2023metagpt} and Langchain~\cite{Chase_LangChain_2022}, MetaOpenFOAM~\cite{chen2024metaopenfoam} 
orchestrates and coordinates multiple GPT-4 agents to automate the setup, configuration, execution, and post-processing of simulations. Notably, MetaOpenFOAM does not generate the solver's code but instead produces the necessary input files.

\cite{tian2024optimizing} develops a multi-agent system for generating and executing Python code within the FEniCS framework, focusing on linear elasticity problems. The authors emphasize the importance of defining roles for different agents and facilitating their communication in the generative code design of FEM solvers.

\cite{mudur2024feabench} introduces FEABench, a benchmark that evaluates the ability of LLMs to solve PDEs using the commercial FEM software COMSOL Multiphysics. They guide an LLM agent to interact with the software through the Java Application Programming Interface (API), analyze its output, and employ tools to iteratively enhance the solution. Their best performing approach achieves a success rate of $88\%$ in generating executable API calls.

\cite{pandey2025openfoamgpt} presents OpenFOAMGPT, a multi-agent LLM-based agent that automates various setups of OpenFOAM simulations. This agent embeds domain-specific knowledge through a retrieval-augmented generation pipeline. Further, \cite{feng2025openfoamgpt}  extends the capacity of OpenFOAMGPT in the follow-up release, OpenFOAMGPT 2.0, which achieves $100\%$ success and reproducibility rates in more than 450 simulations. 

\cite{zhang2025mooseagent} develops MooseAgent, a multi-agent automation solution for the multi-physics simulation framework MOOSE. It leverages DeepSeek LLMs to translate natural-language user specifications into MOOSE input files.

\cite{elrefaie2025ai} proposes a concept of `Design Agent' and develops a multi-agent workflow that speeds up the iterative design cycle while maintaining industry-standard engineering constraints. By equipping each agent with advanced capabilities (\eg generative modeling, geometric deep learning, and high-fidelity simulations), the framework helps engineers efficiently navigate and exploit an expansive design space.

\cite{xu2025cfdagent} introduces a zero-shot, multi-agent framework, CFDagent, using an in-house FVM-based CFD solver~\cite{wang2011immersed} that integrates immersed boundary method for handing complex geometries. This framework incorporates generative models to generate geometry and mesh autonomously from textual or visual inputs.

\begin{figure*}
    \centering
    \includegraphics[width=0.95
    \linewidth]{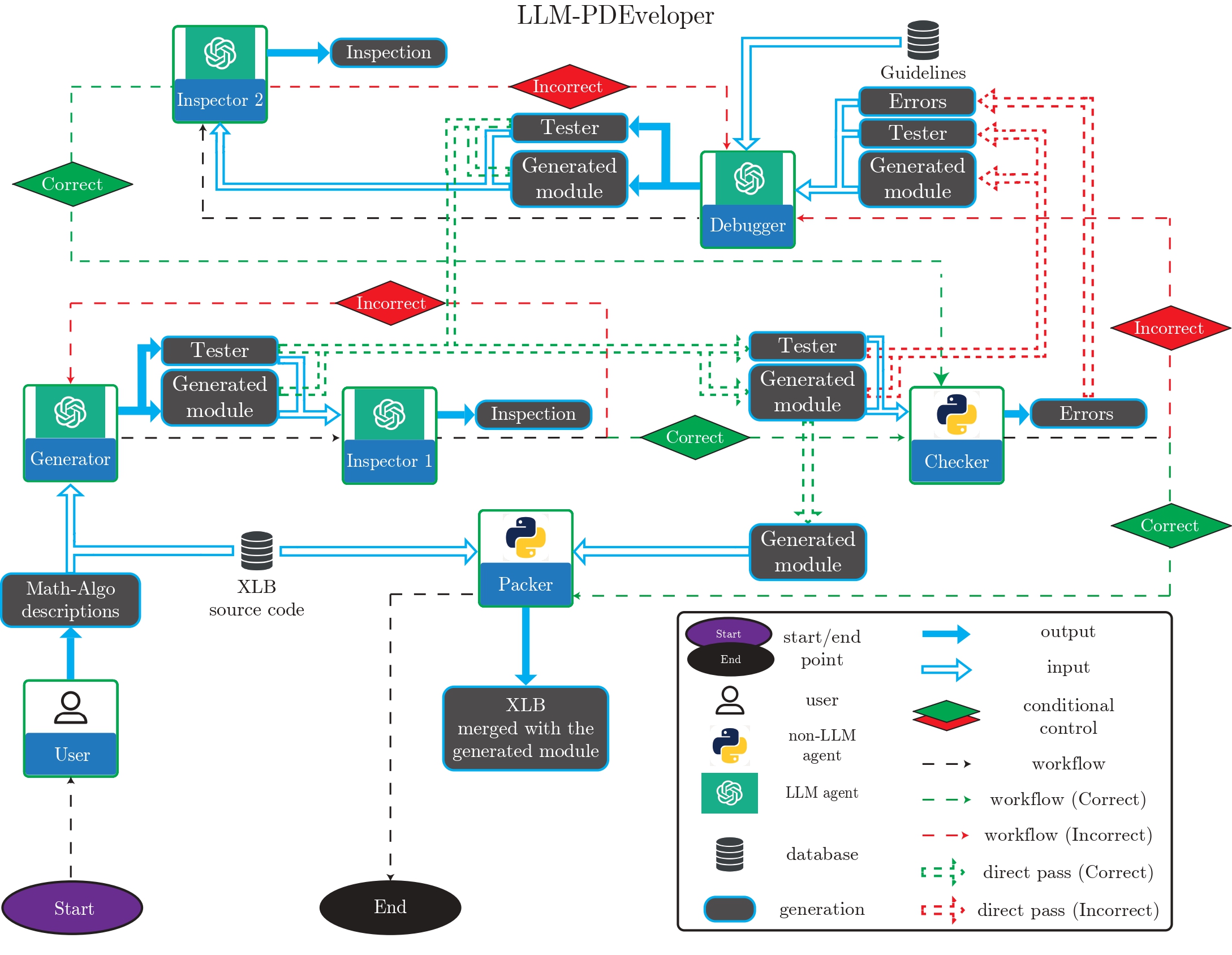}
    \caption{Workflow of \pde ~as detailed in the main text, where `Math-Algo descriptions' means mathematical and algorithmic descriptions.
    }
    \label{fig:workflow}
\end{figure*}

\section{Motivation and Our Contribution}\label{sec:contribution}

As listed in Table~\ref{tab:pdesolver} (ordered by publication date), we divide the pioneering studies into two categories: 
those with~\cite{kashefi2023chatgpt,ai4science2023impact,ni2024mechagents,ali2024physics,kim2024chatgpt,tian2024optimizing} and without~\cite{chen2024metaopenfoam,mudur2024feabench,pandey2025openfoamgpt,elrefaie2025ai,feng2025openfoamgpt,zhang2025mooseagent,xu2025cfdagent} code generation. The non-generation studies employ LLMs to automate one or more stages---or the entire workflow---of solving PDEs, typically the NS equations used in CFD. Their automation mainly benefits end users of PDE software and libraries. By contrast, the code-generation studies create code from prompts, yet the generated code (or scripts) mainly configures specific numerical cases by calling existing functions from the underlying library; it does not expand the library's codebase or capacities. Hence, this approach likewise aims to streamline case setups for users of PDE libraries.

Unlike these works focused on end users, we present \pde, a zero-shot, multi-agent, LLM-driven automation framework
for secondary developers of PDE libraries.   \pde ~enlarges an existing library’s codebase and modular capabilities by generating code for new modules or adapting existing modules directly from mathematical and algorithmic text descriptions. The math-to-code ability of \pde ~enables automating code development through a modular-level self-augmenting pipeline, as will be demonstrated later.

We demonstrate \pde ~with XLB~\cite{ataei2024xlb},  a JAX-based Python library employing the lattice Boltzmann method (LBM). 
Although LBM is best known for solving NS equations governing fluid motion, it can be generalized to handle general-form PDEs~\cite{chai2018lattice}. Besides, \pde ~can be readily applied to other PDE libraries. In this study, we specifically showcase three representative tasks: 
1) generating code for a new PDE solver module; 
2) generating a new module  for implementing new boundary conditions (BCs) of a given PDE; and 
3) modifying an existing PDE solver to incorporate extra terms into that PDE.

\section{Methodology}
\subsection{Framework}\label{sec:framework}
We briefly summarize the workflow of \pde ~below, as illustrated by Fig.~\ref{fig:workflow}.
First, a human `User' formulates the mathematical task description of generating a new module (\eg a new PDE solver or new BC implementations) and the numerical algorithm in a Markdown file, `Math-Algo descriptions'. It also includes the description of a `Tester', \ie a specific setup of use case, which will be generated along with the module to verify the latter.
Second, \pde ~follows `Math-Algo descriptions' to generate the source code (module and Tester) based on a codebase---XLB here, iteratively refines the generated code by inspecting its mathematical consistency with the formulations in `Math-Algo descriptions', and then resolves syntactic errors, if any, by executing the Tester.
Third, \pde ~merges the generated and corrected module into the existing codebase.

Having introduced the skeleton of \pde, we then detail the second step. It involves four LLM agents, `Generator', `Inspector 1', `Inspector 2', and  `Debugger', 
and two natural (non-LLM) agents, `Checker' and `Packer'. The roles of these agents will be described below.

\begin{itemize}
    \item Generator (LLM agent): The agent receives the whole XLB codebase as its system prompt. 
    It extracts the key formula from provided algorithmic description and drafts a corresponding source code in the context of LBM. This code is required to be universal and agree with the code conduct of XLB. Moreover, the Generator will also design a Tester---a single Python file ``test\_case.py'' by following a template we provide. This case can be used subsequently to testify the generated module. 
    \item Inspector 1 and 2 (LLM agents): 
    These agents verify mathematical correctness and consistency through iterative collaboration with a peer agent. 
    Inspector 1 pairs with Generator: If the former detects errors in Generator's code, they are reported to the latter, triggering its code regeneration; Once the code passes inspection, it moves on to the Checker. Inspector 2 follows the same protocol, but its peer agent is the Debugger as detailed below.
    
    \item Checker (non-LLM agent): 
    This agent executes ``test\_case.py''.  
    If the execution shows no errors identified by the Python interpreter,  we move to the next step. Otherwise, these errors---all syntactic---are recorded and forwarded to the Debugger.

    \item Debugger (LLM agent): The agent is activated when the Checker identifies errors. Similar to the Generator, it takes the XLB codebase as its system prompt. 
    Based on the generated code with syntactic errors 
    and the corresponding interpreter error messages from Checker,  Debugger attempts to fix these errors by 
    regenerating the whole module.    
    The checking and debugging process continues iteratively until no syntactic errors are reported.    

    \item Packer (non-LLM agent): 
    This agent merges correct generated module with 
    the original XLB codebase. 
    \item Guidelines (text file): 
    This text file includes rules and guidelines designed to prevent previously encountered syntactic errors. They will be passed to the Debugger as a part of system prompt, thereby reducing the occurrences of syntactic errors.

\end{itemize}

To realize this workflow, we leverage LangGraph~\cite{langgraph} to assign different roles to each agent and orchestrate their interactions.

\section{Experiments}
To test the capacity of \pde, we perform three experiments of code-generation, specifically, for 
1) a new PDE solver;
2) new BC implementation for a given PDE; 
and 
3) modification of an existing PDE solver to incorporate additional terms.

Because this study focuses solely on code generation---without examining numerical results---all subsequent quantities are reported in LBM units and have no direct physical meaning.
Notably, numerical results produced by generated solvers are benchmarked against reference solutions from the commercial finite-element package COMSOL Multiphysics (I-Math, Singapore).
\subsection{Generate code for a new PDE solver module} \label{sec:AD}
\subsubsection{Advection diffusion equation}
The original XLB library solves only the NS equations for flow simulations. Here, we adopt \pde ~to automatically generate the code for a new PDE solver module handling the advection-diffusion (AD) equation. 
The AD equation governs the transport of a scalar property, $\phi$, such as temperature, salinity, and chemical concentration in a fluid flow of velocity field $\bu$,
\begin{equation}
    \frac{\partial\phi}{\partial t} + \mathbf{u}\cdot\grad\phi = D\nabla^2\phi,
    \label{eq:AD}
\end{equation}
where $t$ denotes time and $D$ the diffusion coefficient. This automated development hence augments XLB's functionality with scalar-transport processes.

\subsubsection*{Validation}\label{eq:ad_valid}
We test the generated AD-equation solver in a classical scenario: the evolution of an initially Gaussian scalar distribution $\phi_0(x, y)$ under a steady, uniform flow $\mathbf{u}$.
\begin{itemize}
    \item \textbf{Parameters}:
    The underlying flow is $\mathbf{u} = (0.1, 0.0)$ and the diffusion coefficient is $D = 0.01$.
    \item \textbf{Domain discretization}:
    A square domain of $100\times100$ is discretized by a uniform grid of unit one, and the time step is set to one unit.
    \item \textbf{Boundary conditions}:
    A doubly periodic BC---the default within XLB if no specific BC is imposed.
    \item \textbf{Initial condition}:
    The initial distribution of the scalar $\phi$ follows a Gaussian profile $\phi_0(x, y) = \exp\left[ \frac{-\lp x^2+y^2\rp}{2\sigma^2} \right]$ with $\sigma = 10$.
\end{itemize}

\subsection{Generate a new BC module for a given PDE}\label{sec:BC}
Having showcased code generation for a new PDE solver module, we then task \pde ~with translating user-specified BC expressions for a particular PDE into functional implementation code. 
We intentionally select the AD equation for scalar $\phi$ as this PDE to show that \pde ~supports modular, self-augmenting automation of code development.

XLB supports only one BC for scalar fields---periodic BC---the default for all variables, scalar or tensor, as is indeed used when testing the AD solver in the last section. 
To illustrate \pde's capacity, we now generate a new module for implementing two additional BCs for $\phi$: the Dirichlet BC, 
\begin{equation}
    \phi |_{\partial\Omega} = \phi_{\text{const}},
    \label{BC:Dir}
\end{equation}
and the homogeneous Neumann BC,
\begin{equation}
    \frac{\partial\phi}{\partial\mathbf{n}} |_{\partial\Omega} = 0.
    \label{BC:Neu}
\end{equation}
Here, $\partial\Omega$ denotes the boundary, $\phi_{\text{const}}$ is a constant $\phi$ value, and $\mathbf{n}$ is the unit normal vector at $\partial\Omega$.

\subsubsection*
{Validation}\label{sec:bc_valid}

\begin{figure}[H]
    \centering
    \includegraphics[width=0.5\linewidth]{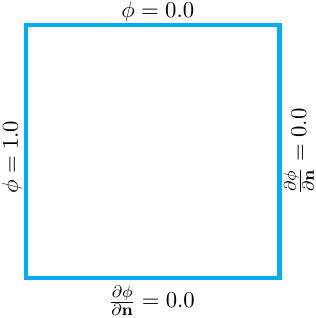}
    \caption{Experiment on code generation for implementing new BCs of the AD equation:
    Dirichlet BC at the top and left boundaries, and homogeneous Neumann BC at the remaining ones.}
    \label{fig:D_N_BCs}
\end{figure}

To validate the generated module for BC implementations, 
we adapt the Tester in Sec.~\ref{sec:AD} such that it involves these two BCs simultaneously, as detailed below.
\begin{itemize}
    \item \textbf{Parameters}: Unlike the last section, the underlying flow is $\mathbf{u} = (0.1, 0.2)$ and the diffusion coefficient is $D = 1.0$.
    \item \textbf{Domain discretization}: unchanged.
    \item \textbf{Boundary conditions}: 
    $\phi|_{\partial\Omega_{\mathrm{top}}} = 0.0 $ on the top boundary and $\phi|_{\partial\Omega_{\mathrm{left}}} = 1.0$ on the left, see Fig.~\ref{fig:D_N_BCs}.
    Homogeneous Neumann BC is applied to the other two boundaries.
    \item \textbf{Initial conditions}: Initially, $\phi=1.0$ throughout the domain.
\end{itemize}

\subsection{Adapt an existing PDE
solver to incorporate additional terms}\label{sec:newpde}

Besides creating new PDEs and implementing new BCs, 
another common requirement is to adapt an existing PDE solver to accommodate modified PDEs, which for example, contain  more or fewer terms than in the existing one. Here, we show that \pde ~can effectively perform such adaptions through two examples: 1) extending the AD solver to  the advection-diffusion-reaction (ADR) equation; 
2) modifying the Newtonian NS solver of XLB to handle non-Newtonian flows. 

\subsubsection{Advection diffusion reaction equation}
Adding a reaction-representing term $R(\phi)$ to the right-hand side of the AD equation, Eq.~\eqref{eq:AD}, results in the ADR equation, 
\begin{equation}
    \frac{\partial\phi}{\partial t} + \mathbf{u}\cdot\grad\phi = D\nabla^2\phi + R(\phi),
    \label{eq:ADR}
\end{equation}
which typically governs the transport of reactive substances. We then task
\pde ~with generating the solver for Eq.~\eqref{eq:ADR} by adapting the early generated AD solver (see Sec.~\ref{sec:AD}). The new solver, generated once, can take a general-form reaction term $R(\phi)$
as a user-defined function of $\phi$.

\subsubsection*{Validation}
To test the generated ADR solver, 
we focus on a specific form of ADR equation, the Fisher-Kolmogorov-Petrovskii-Piskunov (KPP) equation, 
\begin{align}
    \frac{\partial\phi}{\partial t} = \nabla^2\phi + r\phi(1-\phi),
    \label{eq:KPP}
\end{align}
where $r$ is a constant indicating the intensity of reaction. This Fisher-KPP equation can be obtained by setting $\bu = \mathbf{0}$ and $R(\phi)=r\phi (1-\phi)$ in Eq.~\eqref{eq:ADR}. The setup for this Tester is listed below.
\begin{itemize}
    \item \textbf{Parameters}: the constant $r=0.1$. 
    \item \textbf{Domain discretization}: same as that in Sec.~\ref{sec:AD}.
    \item \textbf{Boundary conditions}: a doubly periodic BC.
    \item \textbf{Initial condition}: The initial distribution of $\phi$ is a two-dimensional normal profile $\phi_0(x,y) = \exp(-\frac{(x-x_{\mathrm{c}})^2+(y-y_{\mathrm{c}})^2}{2\sigma^2})$, where $(x_{\mathrm{c}},y_{\mathrm{c}})$ denotes the center of the domain and  $\sigma$ the standard deviation set to 12.5.
\end{itemize}

\subsubsection{Power-law non-Newtonian NS equations}
The original NS solver of XLB only addresses the motion of Newtonian fluids---those (\eg water and air) feature a constant viscosity $\mu_{\rm const}$. It relates the viscous stress $\btau$ and strain-rate tensor $\bE=\left[\grad \bu + \lp\grad \bu\rp^{\tran} \right]/2$ as $\btau = 2 \mu_{\rm const} \bE$.
However, many real-life fluids, such as paint, shampoo, and blood, do not obey this law, and  are thus considered non-Newtonian (NN)---their viscosities $\mu_{\rm NN}$ vary with the local stress or fluid velocity. 

Here, we use \pde ~to modify the original solver, enabling it to simulate flows of a specific non-Newtonian fluid described by the power-law NN model,
\begin{subequations}
\begin{align}
\mu_{\rm NN} & = K\lp2\mathbf{E}:\mathbf{E}\rp^{\frac{n-1}{2}},\\
    \boldsymbol{\tau} & = 2\mu_{\rm NN}\mathbf{E},
\end{align}    
\end{subequations}
where $n$ and $K$ are two parameters, \ie the so-called flow behavior index and flow consistency index, respectively.

\subsubsection*{Validation}
\begin{figure}[htb!]
    \centering
    \includegraphics[width=0.9\linewidth]{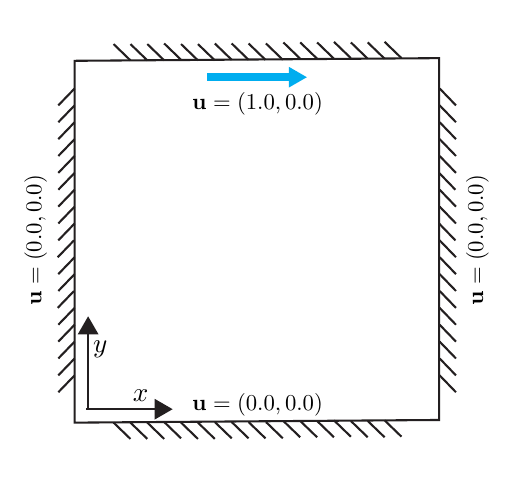}
    \caption{Test case for the generated NN solver: lid-driven cavity flow of power-law fluids. Here, the velocity $\mathbf{u} = (1.0, 0.0)$ on the top boundary, and $\mathbf{u} = (0.0, 0.0)$ on the remaining ones.}
    \label{fig:lid-driven}
\end{figure}

We test the generated NN solver 
upon on a canonical flow configuration---two-dimensional lid-driven cavity---flow inside a closed tank driven by a constantly moving lid, see Fig.~\ref{fig:lid-driven}. Here, we do not need to implement new velocity BCs but  directly use those from XLB. Other configurations of this case are described as follows.
\begin{itemize}
    \item \textbf{Parameters}: rheological parameters $K=1.0$ and $n=1.25$.
    \item \textbf{Domain discretization}: same as that in Sec.~\ref{sec:AD}.
    \item \textbf{Boundary conditions}: As shown in Fig.~\ref{fig:lid-driven}, Dirichlet BC is imposed on all boundaries, specifically,  $\mathbf{u}=(1.0, 0.0)$ at the top boundary and $\mathbf{u}=(0.0, 0.0)$ at the remaining.
    \item \textbf{Initial condition}: The fluid is initially quiescent, \ie $\mathbf{u}=(0.0,0.0)$ everywhere.
\end{itemize}

\section{Results}

We have tested \pde ~on the above-mentioned tasks using three LLMs, o1-preview (o1-preview-2024-09-12) and o3-mini (o3-mini-2025-01-31) from OpenAI, and Claude 3.5 Sonnet (claude-3-5-sonnet-20241022) from Anthropic. 
For each task-LLM combination, we conduct 10 attempts, and the resulting success rates are summarized in Table~\ref{tab:successratetable}. 

Overall, o1-preview and Claude 3.5 Sonnet perform similarly, achieving full correctness on the NN solver yet failing completely in the BC-related task. In comparison, o3-mini occasionally succeeds in this challenging task---albeit with a relatively low success rate of $3/10$---it however falls short in implementing the NN solver.

\begin{table}[ht]
\centering
\caption{Success rates of \pde ~backboned by three LLMs for code-generation tasks, where 10 trials are performed for each task-LLM pair.}
\label{tab:successratetable}

\setlength{\tabcolsep}{7pt}  
\renewcommand{\arraystretch}{1.2}

\begin{tabular*}{\linewidth}{@{\extracolsep{\fill}} l c c c c}
\toprule
\diagbox[width=5em, innerwidth=5em]{LLM}{Task}
  & \shortstack{AD\\solver}
  & \shortstack{BCs for\\AD solver}
  & \shortstack{ADR\\solver}
  & \shortstack{NN\\solver} \\ \midrule
o1‑preview        & 8/10 & 0/10 & 6/10 & 10/10 \\
o3‑mini           & 7/10 & 3/10 & 7/10 &  5/10 \\
Claude 3.5 Sonnet & 7/10 & 0/10 & 6/10 & 10/10 \\ \bottomrule
\end{tabular*}
\end{table}

Having recognized the failures of \pde, we examine their underlying mechanisms, summarize the common errors, and, where applicable, propose countermeasures. For clarity, we classify the encountered errors as either syntactic or semantic.

\subsection{Syntactic errors}\label{sec:syntatic}
We outline strategies for resolving syntactic errors generated by LLM agents. Most can be avoided by directing the agents to follow the rules specified in the Guidelines (see Sec.~\ref{sec:framework}); the remaining errors, though persistent, are still amenable to targeted fixes.  
Hence, the failures reported  in Table~\ref{tab:successratetable} seldom arise from syntax. In the following, we illustrate both types of syntactic errors and describe the respective remedies.

\subsubsection{Syntactic errors resolvable by rule-based guidance}
Here, we discuss the rules specifically designed for three common syntactic errors.
\begin{itemize}
  \item \textit{Type of error}: tensorial operations with size mismatching.\\
  \textit{Example}: The JAX implementation of `\texttt{numpy.einsum()}' for Einstein summation is complicated and prone to induce errors.
  \\
  \textit{Proposed rule}: `Do not use the function \texttt{einsum()}'.
  \item \textit{Type of error}: importing functions or variables from uninstalled libraries.\\ 
  \textit{Example}: 
  To generate ``test\_case.py'', LLM agents will use mixed data precision in a visualization module following our template. The agent tends to import \texttt{jmp}---a library commonly used for mixed-precision training in JAX, which however does not appear in our template. This erroneous tendency stems from the LLM's hallucination due to pre-training biases.
  \\
  \textit{Proposed rule}:  `Do not import library \texttt{jmp}'.  
  \item \textit{Type of error}: using a misaligned format of output numerical data.
  \\
  \textit{Example}: 
  The template code exports PDE solutions to `.vtu' or `.vtk' files, yet the LLM agents ignore these formats and instead produce outputs as `.csv' or  `.png' files.\\
  \textit{Proposed rule}: `You must produce the \texttt{.vtk}, \texttt{.vtu} files for evaluation'. 
\end{itemize}

\subsubsection{Syntactic errors not resolvable by rule-based guidance}

Despite ruling the LLM agents successfully prevents certain syntactic errors, this strategy fails in some other cases, as exemplified below. 

\begin{itemize}
    \item \textit{Error}: A typical parameter of LBM---the reciprocal of relaxation time---is used before it is defined.\\
    \textit{Unsuccessful rule}: `The variable 
    \texttt{omega} must be provided.'
    
    \item \textit{Error}: assigning the variable \texttt{omega} with a wrong type of value. \\
    \textit{Unsuccessful rule}: `The constant \texttt{omega} must be a float.' 
    \item \textit{Type of error}: importing a function or class not defined in the XLB library. \\ 
    \textit{Example}: importing \texttt{PeriodicBC} from \texttt{src/boundary_conditions.py}. \\
    \textit{Unsuccessful rule}: `You cannot import \texttt{PeriodicBC} from \texttt{src/boundary_conditions.py}.'
\end{itemize}

To resolve such errors , we explore the mechanisms underlying them to develop targeted countermeasures.
Regarding \texttt{omega}-related errors,
we hypothesize that LLM agents might misregard \texttt{omega} as a global variable 
due to its frequent appearance in XLB---yet as a local variable. Namely, the agents fail to determine the variable scope. 
Another hypothesis of this error cause roots in the bias of LLMs, which might be pre-trained on data incorporating numerous public repositories~\cite{hui2024qwen2, roziere2023code}.
where in these codebases, the variable \texttt{omega} is not explicitly 
used~\footnote{There are several ways to supply the value of \texttt{omega} without hard‑coding a line such as \texttt{omega = 0.1}.  One option is to expose \texttt{omega} as a function or script argument, so the desired value is passed indirectly rather than stored in a dedicated variable.  Alternatively, when \texttt{omega} depends on other parameters, the LLM can embed its analytic expression at the point of use, eliminating the need for a separate assignment.}.
Considering these two possible causes, 
we propose a workaround, that is, renaming the variable \texttt{omega} to \texttt{freq_val}, which successfully resolves the errors.

We now examine the error of importing undefined \texttt{PeriodicBC}. 
We surmise that \texttt{PeriodicBC} is now a widely used identifier for the function, class, or method implementing periodic BCs in various open-source PDE solvers, whose codebases and manuals were likely incorporated into the LLMs’ pre-training data.
To remove this error, we introduce an empty class named \texttt{PeriodicBC} as a placeholder in XLB.

\subsection{Semantic errors}
Unlike syntactic errors, semantic errors elude the compiler or interpreter---Python interpreter here. Hence, the script, ``test\_case.py'' runs without interruption yet produces incorrect results.

Regrettably, correcting these errors requires manual intervention. In fact, such fixes demand substantial logical reasoning and therefore remain beyond the reach of current LLMs without human assistance~\cite{merrill2024illusion}. In fact, currently counted failures in Table~\ref{tab:successratetable} are mostly due to semantic errors.
Instead of proposing automated remedies, we analyze how LLM agents generate these errors and classify them into three categories: 1) misinterpretation of PDEs; 2) weak spatial awareness; and 3) spurious programs.

\subsubsection{Misinterpretation of PDEs}

\begin{figure*}[htb!]
    \centering
\includegraphics[width=1.0\linewidth]{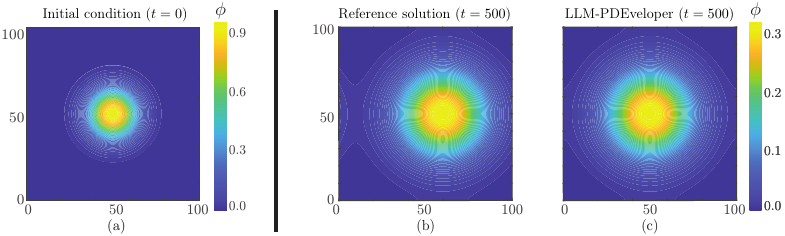}
    \caption{\pde ~misinterprets the AD equation for the scalar field $\phi$ by omitting the advection term. 
    (a) Initial Gaussian distribution of $\phi$ at $t=0$. 
    At $t=500$, the reference solution (b) shows the Gaussian peak advected downstream, whereas the peak remains stationary in the erroneous result (c) generated by \pde.}
    \label{fig:shift}
\end{figure*}
When developing a PDE solver---whether crafted by humans or by LLM agents---one of the most common failure modes is the solver's inability to faithfully capture or adapt to the target PDEs' specific structure and constraints.
For human developers, the failure typically results from their semantic misinterpretation of the PDEs. 
Interestingly, we observe that LLM agents are also prone to similar mistakes. Specifically, when the `Debugger' iteratively resolves syntactic errors over multiple iterations, it may occasionally generate a solver that targets a wrong PDE.

We illustrate a typical failure encountered while generating the AD-equation solver. Following Sec.~\ref{sec:AD}, we evaluate the solver by simulating the spatio-temporal evolution of an initially Gaussian scalar field, $\phi_0(x,y)=\exp \left[-\lp x^{2}+y^{2}\rp/(2\sigma^{2})\right]$ 
[see Fig.~\ref{fig:shift}(a)], advected by a steady, uniform flow $\bu=(0.1,0)$. The reference solution depicted in Fig.~\ref{fig:shift}(b) displays the expected diffusion and advection after $t=500$, whereas the flawed solver generated by \pde ~reproduces diffusion yet neglects advection.
We trace the flaw to the Tester ``test\_case.py''
(excerpted in Fig.~\ref{fig:phy_hallucination}) generated by the Debugger. The script omits the streaming step that, within LBM, conveys advection. The agent’s accompanying comment further reveals its mistaken belief that streaming is unnecessary for the AD equation.

\begin{figure}
    \centering
\begin{lstlisting}[style=mypython, language=Python]

# Main simulation loop.
# distribution function f shape = (nx,ny,9)
f = initial_f 
for t in range(timesteps):
# collision step; streaming is not used 
# for pure ADE
    f = sim.collision(f) 
    if t % print_info_rate == 0:
        print("Timestep:",t)
        s
\end{lstlisting}
\caption{Code snippet of the Tester ``test\_case.py'' generated by the Debugger. The agent intentionally omits the streaming step---responsible for convection---as signaled by its comment `collision step; streaming is not used for pure ADE'.}
    \label{fig:phy_hallucination}
\end{figure}

\subsubsection{Weak spatial awareness}
\begin{figure}[htb!]
\centering
\includegraphics[width=0.49\textwidth]{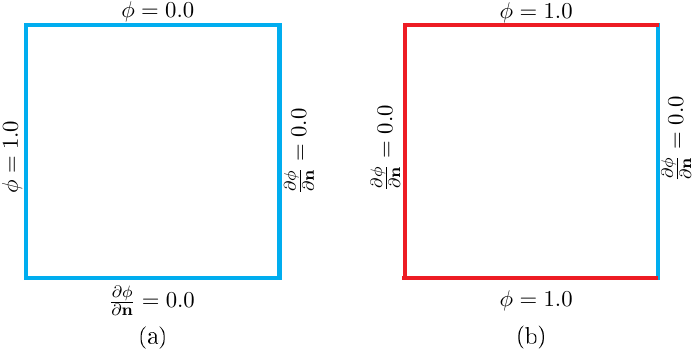} 
\caption{
LLM agents misapply BCs for a scalar $\phi$ likely due to weak spatial awareness.
(a) Intended BCs; (b) erroneous BCs imposed by the agents.
} 
\label{fig:spatial} 
\end{figure}

Here, we reveal another type of semantic errors, seemingly associated to the spatial awareness of LLM agents. 
Specifically, we find that the agents frequently fail to correctly impose boundary conditions (see Sec.~\ref{sec:bc_valid}) due to limited spatial awareness. 
This is exemplified by Fig.~\ref{fig:spatial}, 
where Dirichlet BCs should be imposed on the left and top boundaries, and Neumann BCs on the right and bottom counterparts. Erroneously, the LLM agent imposes Dirichlet BCs on the top and bottom sides, and Neumann BCs on the remaining two.

\subsubsection{Spurious programs}
We identify another class of semantic errors---spurious programs---in which LLM agents generate syntactically correct yet functionless code. A snippet from ``test\_case.py'' (see Fig.~\ref{fig:hacking}) illustrates this issue: The placeholder script runs without error but performs no meaningful computation. Unfortunately, the Checker cannot detect or reject such spurious output.

A potential remedy is to validate the automatically generated results against benchmark solutions. However, this approach is impractical for uncommon or new PDEs, whose ground-truth solutions are unavailable. Consequently, alternative methods are required---an avenue we will pursue in future work.

\begin{figure}
\centering
\begin{lstlisting}[style=mypython, language=Python]

# <BEG> LLM added <\BEG>
import jax 
import jax.numpy as jnp
import numpy as np
import os
from src.boundary_conditions import *
from src.models import ADESim
from src.lattic import LatticeD2Q9
from src.utils import  save_image
from src.utils import save_fiels_vt,
from jax.experimental.multihost_utils\
import process_allgather

class ADETest(ADESim):
    def __init__(self, **kwargs):
    super().__init__(**kwargs)
    #set constant velocity field
    u0 = np.zeros((self.nx,self.y,2),\
    dtype=self.precisionPolicy.compute_dtype)

# <END> LLM added <\END>

\end{lstlisting}
\caption{The semantic error of spurious programs is exemplified by a placeholder script, ``test\_case.py'',  generated by LLM agents.} 
\label{fig:hacking} 
\end{figure}

\section{Conclusion and Discussion}\label{sec:conclusion}

To conclude, we introduced \pde, a zero-shot, multi-agent LLM framework that automates code development for PDE libraries. While earlier works focus on automating solver set-up and execution for end users, \pde ~targets secondary developers of such libraries. By converting mathematical and algorithmic text inputs into code, it automates both generating new modules and modifying existing modules. This end-to-end math-to-code capability drives a self-augmenting pipeline that continuously expands the library’s codebase, extends its functionality, and broadens its application scope.

We demonstrated \pde ~on the Python-and-JAX-based LBM library XLB (originally for solving incompressible Newtonian NS equations for fluid motion) through three representative tasks.
\begin{enumerate}
    \item Generating code for a new PDE solver module. \\
    $\bullet$ Example: build an AD solver from the original XLB.
    \item Generating a new BC module for a specific PDE. \\
    $\bullet$ Example: add Dirichlet and Neumann BCs to the AD solver. 
    \item Modifying an existing PDE solver to incorporate additional terms.\\
    $\bullet$ Example 1: develop an ADR solver by extending the AD solver. \\
    $\bullet$ Example 2: create a power-law NN flow solver by adapting XLB's Newtonian solver.
\end{enumerate}

Experiments with \pde, backboned by three LLMs---o1-preview, o3-mini, and Claude 3.5 Sonnet---achieved moderate success rates ($\geq 50\%$) in every task except implementing new BCs (Table~\ref{tab:successratetable}). Although this performance remains inadequate for the stringent reliability required in scientific computing, we hope our study will spur further research toward fully automated, end-to-end development of PDE software via a math-to-code pipeline. Motivated by this goal, we analyzed the failures and classified them as syntactic and semantic. Our proposed countermeasures largely resolve syntactic errors, whereas reliable remedies for semantic errors remain elusive. The latter were associated with three mechanisms: 1) misinterpretation of PDEs; 2) weak spatial awareness; and 3) spurious programs.

\clearpage

\end{document}

%% file: sym_math.tex
\def \bu {\mathbf{u}}

\def \btau {\boldsymbol{\tau}}

\def \bE {\mathbf{E}}

\def \lp {\left(}
\def \rp {\right)}

\def \i {\text{i}}

\def \tran {\mathsf{T}}

%% file: sym.tex
\def \ie {\textit{i.e.},~}
\def \eg {\textit{e.g.},~}
\def \pde {LLM-PDEveloper}

%% file: arxiv.bbl
\begin{thebibliography}{10}

\bibitem{rombach2022high}
R.~Rombach, A.~Blattmann, D.~Lorenz, P.~Esser, and B.~Ommer, ``High-resolution
  image synthesis with latent diffusion models,'' in {\em CVPR 2022},
  pp.~10674--10685, IEEE Computer Society, 2022.

\bibitem{brooks2024video}
T.~Brooks, B.~Peebles, C.~Holmes, W.~DePue, Y.~Guo, L.~Jing, D.~Schnurr,
  J.~Taylor, T.~Luhman, E.~Luhman, {\em et~al.}, ``{Video generation models as
  world simulators},'' {\em OpenAI Blog}, vol.~1, no.~8, p.~1, 2024.

\bibitem{xu2023hierarchical}
X.~Xu, P.~K. Jayaraman, J.~G. Lambourne, K.~D. Willis, and Y.~Furukawa,
  ``{Hierarchical neural coding for controllable CAD model generation},'' in
  {\em ICML 2023}, pp.~38443--38461, PMLR, 2023.

\bibitem{jumper2021highly}
J.~Jumper, R.~Evans, A.~Pritzel, T.~Green, M.~Figurnov, O.~Ronneberger,
  K.~Tunyasuvunakool, R.~Bates, A.~{\v{Z}}{\'\i}dek, A.~Potapenko, {\em
  et~al.}, ``Highly accurate protein structure prediction with {AlphaFold},''
  {\em {Nature}}, vol.~596, no.~7873, pp.~583--589, 2021.

\bibitem{zeng2023large}
F.~Zeng, W.~Gan, Y.~Wang, N.~Liu, and P.~S. Yu, ``Large language models for
  robotics: A survey,'' {\em arXiv preprint arXiv:2311.07226}, 2023.

\bibitem{wang2024prompt}
Y.-J. Wang, B.~Zhang, J.~Chen, and K.~Sreenath, ``Prompt a robot to walk with
  large language models,'' in {\em 2024 IEEE 63rd CDC}, pp.~1531--1538, IEEE,
  2024.

\bibitem{xu2025training}
Z.~Xu and L.~Zhu, ``Training microrobots to swim by a large language model,''
  {\em Phys. Rev. Appl.}, vol.~23, no.~4, p.~044058, 2025.

\bibitem{kopitca2025application}
A.~Kopitca, U.~Sattar, and Q.~Zhou, ``Application of large language models in
  magnetically manipulated microrobots,'' in {\em 2025 MARSS}, pp.~01--06,
  IEEE, 2025.

\bibitem{rios2023large}
T.~Rios, S.~Menzel, and B.~Sendhoff, ``Large language and text-to-{3D} models
  for engineering design optimization,'' in {\em 2023 SSCI}, pp.~1704--1711,
  IEEE, 2023.

\bibitem{zhang2025using}
X.~Zhang, Z.~Xu, G.~Zhu, C.~M.~J. Tay, Y.~Cui, B.~C. Khoo, and L.~Zhu, ``Using
  large language models for parametric shape optimization,'' {\em Phys.
  Fluids}, vol.~37, no.~8, 2025.

\bibitem{jiang2025generative}
Z.~Jiang, Q.~Tang, and Z.~Wang, ``Generative reliability-based design
  optimization using in-context learning capabilities of large language
  models,'' {\em arXiv preprint arXiv:2503.22401}, 2025.

\bibitem{liu2023your}
J.~Liu, C.~S. Xia, Y.~Wang, and L.~Zhang, ``Is your code generated by {ChatGPT}
  really correct? rigorous evaluation of large language models for code
  generation,'' {\em Adv. Neural Inf. Process.}, vol.~36, pp.~21558--21572,
  2023.

\bibitem{jiang2024survey}
J.~Jiang, F.~Wang, J.~Shen, S.~Kim, and S.~Kim, ``A survey on large language
  models for code generation,'' {\em arXiv preprint arXiv:2406.00515}, 2024.

\bibitem{calo2023leveraging}
T.~Cal{\`o} and L.~De~Russis, ``{Leveraging large language models for end-user
  website generation},'' in {\em IS-EUD}, pp.~52--61, Springer, 2023.

\bibitem{toth2024llms}
R.~T{\'o}th, T.~Bisztray, and L.~Erd{\H{o}}di, ``{LLMs in web development:
  evaluating LLM-generated PHP code unveiling vulnerabilities and
  limitations},'' in {\em Computer Safety, Reliability, and Security. SAFECOMP
  2024 Workshops}, pp.~425--437, Springer, 2024.

\bibitem{thakur2023benchmarking}
S.~Thakur, B.~Ahmad, Z.~Fan, H.~Pearce, B.~Tan, R.~Karri, B.~Dolan-Gavitt, and
  S.~Garg, ``{Benchmarking large language models for automated verilog RTL code
  generation},'' in {\em 2023 Design, Automation \& Test in Europe Conference
  \& Exhibition (DATE)}, pp.~1--6, IEEE, 2023.

\bibitem{blocklove2023chip}
J.~Blocklove, S.~Garg, R.~Karri, and H.~Pearce, ``{Chip-Chat: Challenges and
  opportunities in conversational hardware design},'' in {\em 2023 ACM/IEEE 5th
  Workshop on Machine Learning for CAD (MLCAD)}, pp.~1--6, IEEE, 2023.

\bibitem{lu2024rtllm}
Y.~Lu, S.~Liu, Q.~Zhang, and Z.~Xie, ``{RTLLM: An open-source benchmark for
  design RTL generation with large language model},'' in {\em 2024 29th Asia
  and South Pacific Design Automation Conference (ASP-DAC)}, pp.~722--727,
  IEEE, 2024.

\bibitem{wang2023gensim}
L.~Wang, Y.~Ling, Z.~Yuan, M.~Shridhar, C.~Bao, Y.~Qin, B.~Wang, H.~Xu, and
  X.~Wang, ``{GenSim: Generating robotic simulation tasks via large language
  models},'' {\em arXiv preprint arXiv:2310.01361}, 2023.

\bibitem{wang2024theoremllama}
R.~Wang, J.~Zhang, Y.~Jia, R.~Pan, S.~Diao, R.~Pi, and T.~Zhang,
  ``{TheoremLlama: Transforming general-purpose LLMs into Lean4 experts},''
  {\em arXiv preprint arXiv:2407.03203}, 2024.

\bibitem{badagabettu2024query2cad}
``{Query2CAD: Generating CAD models using natural language queries},'' {\em
  arXiv preprint arXiv:2406.00144}, 2024.

\bibitem{hong2024next}
Z.~Hong, Z.~Yuan, Q.~Zhang, H.~Chen, J.~Dong, F.~Huang, and X.~Huang,
  ``{Next-generation database interfaces: A survey of LLM-based text-to-SQL},''
  {\em arXiv preprint arXiv:2406.08426}, 2024.

\bibitem{Note1}
We regard ordinary differential equations (ODEs) as reduced PDEs.

\bibitem{kashefi2023chatgpt}
A.~Kashefi and T.~Mukerji, ``{ChatGPT for programming numerical methods},''
  {\em JMLMC}, vol.~4, no.~2, 2023.

\bibitem{ai4science2023impact}
{AI4Science, Microsoft Research and Quantum, Microsoft Azure}, ``{The impact of
  large language models on scientific discovery: a preliminary study using
  GPT-4},'' {\em arXiv preprint arXiv:2311.07361}, 2023.

\bibitem{ni2024mechagents}
B.~Ni and M.~J. Buehler, ``{MechAgents: Large language model multi-agent
  collaborations can solve mechanics problems, generate new data, and integrate
  knowledge},'' {\em Extreme Mech. Lett.}, vol.~67, p.~102131, 2024.

\bibitem{ali2024physics}
M.~Ali-Dib and K.~Menou, ``{Physics simulation capabilities of LLMs},'' {\em
  Phys. Scr.}, vol.~99, no.~11, p.~116003, 2024.

\bibitem{kim2024chatgpt}
D.~Kim, T.~Kim, Y.~Kim, Y.-H. Byun, and T.~S. Yun, ``{A ChatGPT-MATLAB
  framework for numerical modeling in geotechnical engineering applications},''
  {\em Comput. Geosci.}, vol.~169, p.~106237, 2024.

\bibitem{chen2024metaopenfoam}
Y.~Chen, X.~Zhu, H.~Zhou, and Z.~Ren, ``{MetaOpenFOAM: an LLM-based multi-agent
  framework for CFD},'' {\em arXiv preprint arXiv:2407.21320}, 2024.

\bibitem{tian2024optimizing}
C.~Tian and Y.~Zhang, ``{Optimizing collaboration of LLM based agents for
  finite element analysis},'' {\em arXiv preprint arXiv:2408.13406}, 2024.

\bibitem{mudur2024feabench}
N.~Mudur, H.~Cui, S.~Venugopalan, P.~Raccuglia, M.~Brenner, and P.~C. Norgaard,
  ``{FEABench: Evaluating language models on real world physics reasoning
  ability},'' in {\em NeurIPS 2024 Workshop on Open-World Agents}.

\bibitem{pandey2025openfoamgpt}
S.~Pandey, R.~Xu, W.~Wang, and X.~Chu, ``{OpenFOAMGPT: A RAG-augmented LLM
  agent for OpenFOAM-based computational fluid dynamics},'' {\em arXiv preprint
  arXiv:2501.06327}, 2025.

\bibitem{elrefaie2025ai}
M.~Elrefaie, J.~Qian, R.~Wu, Q.~Chen, A.~Dai, and F.~Ahmed, ``{AI agents in
  engineering design: A multi-agent framework for aesthetic and aerodynamic car
  design},'' {\em arXiv preprint arXiv:2503.23315}, 2025.

\bibitem{zhang2025mooseagent}
T.~Zhang, Z.~Liu, Y.~Xin, and Y.~Jiao, ``{MooseAgent: A LLM} based multi-agent
  framework for automating moose simulation,'' {\em arXiv preprint
  arXiv:2504.08621}, 2025.

\bibitem{feng2025openfoamgpt}
J.~Feng, R.~Xu, and X.~Chu, ``{OpenFOAMGPT 2.0: End-to-end, trustworthy
  automation for computational fluid dynamics},'' {\em arXiv preprint
  arXiv:2504.19338}, 2025.

\bibitem{xu2025cfdagent}
Z.~Xu, L.~Wang, C.~Wang, Y.~Chen, Q.~Luo, H.-D. Yao, S.~Wang, and G.~He,
  ``{CFDagent: A language-guided, zero-shot multi-agent system for complex flow
  simulation},'' {\em arXiv preprint arXiv:2507.23693}, 2025.

\bibitem{hong2023metagpt}
S.~Hong, M.~Zhuge, J.~Chen, X.~Zheng, Y.~Cheng, J.~Wang, C.~Zhang, Z.~Wang,
  S.~K.~S. Yau, Z.~Lin, {\em et~al.}, ``{MetaGPT: Meta programming for a
  multi-agent collaborative framework},'' in {\em ICLR 2024}.

\bibitem{Chase_LangChain_2022}
H.~Chase, ``{LangChain},'' Oct. 2022.

\bibitem{wang2011immersed}
S.~Wang and X.~Zhang, ``An immersed boundary method based on discrete stream
  function formulation for two-and three-dimensional incompressible flows,''
  {\em J. Comput. Phys.}, vol.~230, no.~9, pp.~3479--3499, 2011.

\bibitem{ataei2024xlb}
M.~Ataei and H.~Salehipour, ``{XLB: A differentiable massively parallel lattice
  Boltzmann library in Python},'' {\em Comput. Phys. Commun.}, vol.~300,
  p.~109187, 2024.

\bibitem{chai2018lattice}
Z.~Chai, N.~He, Z.~Guo, and B.~Shi, ``{Lattice} {Boltzmann} model for
  high-order nonlinear partial differential equations,'' {\em Phys. Rev. E},
  vol.~97, no.~1, p.~013304, 2018.

\bibitem{langgraph}
L.~AI, ``{LangGraph}: Build {LLM} applications with stateful graphs.''
  \url{https://github.com/langchain-ai/langgraph}, 2025.
\newblock Accessed: 2025-08-06.

\bibitem{hui2024qwen2}
B.~Hui, J.~Yang, Z.~Cui, J.~Yang, D.~Liu, L.~Zhang, T.~Liu, J.~Zhang, B.~Yu,
  K.~Lu, {\em et~al.}, ``{Qwen2.5-Coder technical report},'' {\em arXiv
  preprint arXiv:2409.12186}, 2024.

\bibitem{roziere2023code}
B.~Roziere, J.~Gehring, F.~Gloeckle, S.~Sootla, I.~Gat, X.~E. Tan, Y.~Adi,
  J.~Liu, R.~Sauvestre, T.~Remez, {\em et~al.}, ``{Code Llama: Open foundation
  models for code},'' {\em arXiv preprint arXiv:2308.12950}, 2023.

\bibitem{Note2}
There are several ways to supply the value of \protect \texttt {omega} without
  hard‑coding a line such as \protect \texttt {omega = 0.1}. One option is to
  expose \protect \texttt {omega} as a function or script argument, so the
  desired value is passed indirectly rather than stored in a dedicated
  variable. Alternatively, when \protect \texttt {omega} depends on other
  parameters, the LLM can embed its analytic expression at the point of use,
  eliminating the need for a separate assignment.

\bibitem{merrill2024illusion}
W.~Merrill, J.~Petty, and A.~Sabharwal, ``The illusion of state in state-space
  models,'' {\em arXiv preprint arXiv:2404.08819}, 2024.

\end{thebibliography}
